\documentclass[12pt]{article}

\usepackage[utf8]{inputenc}
\usepackage{times}

\usepackage{graphicx}
\usepackage{dcolumn}
\usepackage{bm}

\usepackage[superscript]{cite}
\usepackage{lineno}
\usepackage{physics}
\usepackage{newtxmath}
\usepackage{upgreek}
\usepackage{gensymb}
\usepackage{units}
\usepackage{float}
\usepackage{multirow}
\usepackage[inkscapelatex=false]{svg}
\usepackage[colorlinks=true,linkcolor=blue,urlcolor=blue,citecolor=blue]{hyperref}

\usepackage[commentmarkup=uwave,deletedmarkup=sout]{changes}
\definechangesauthor[name=Malte Grosche, color=blue]{fmg}

\makeatletter
\renewcommand\@biblabel[1]{#1.}
\makeatother

\usepackage{gensymb}
\usepackage{mathtools}
\usepackage{caption}
\newenvironment{sciabstract}{%
\begin{quote} \bf}
{\end{quote}}

\title{\vspace{-1cm} Electrically-controllable superconducting\\memory effect in UTe$_2$}

\author{Zheyu Wu,$^{1}$ Hanyi Chen,$^{1}$ Mengmeng Long,$^{1}$ Daniel Shaffer,$^{2}$\\Dmitry V. Chichinadze,$^{3,4}$  Andrej Cabala,$^{5}$ Theodore I. Weinberger,$^{1}$\\Alexander J. Hickey,$^{1}$ Jinxu Pu,$^{6}$ Dave Graf,$^{3}$ Vladimir Sechovský,$^{5}$\\Michal Vali{\v{s}}ka,$^{5}$ Gang Li,$^{7,8}$ Rui Zhou,$^{7,8}$\\F. Malte Grosche,$^{1}$ Alexander G. Eaton$^{1\ast}$\\
\\
\normalsize{$^{1}$Cavendish Laboratory, University of Cambridge,}\\
\normalsize{JJ Thomson Avenue, Cambridge, CB3 0US, UK}\\
\normalsize{$^{2}$Department of Physics, University of Wisconsin-Madison,}\\
\normalsize{Madison, Wisconsin 53706, USA}\\
\normalsize{$^{3}$National High Magnetic Field Laboratory, Tallahassee, Florida 32310, USA}\\
\normalsize{$^{4}$Department of Physics, Washington University in St. Louis, St. Louis, MO, 63130 USA}\\
\normalsize{$^{5}$Charles University, Faculty of Mathematics and Physics, Department of}\\
\normalsize{ Condensed Matter Physics, Ke Karlovu 5, Prague 2, 121 16, Czech Republic}\\
\normalsize{$^{6}$School of Physics and Astronomy, Shanghai Jiao Tong University, Shanghai, 200240, China}\\
\normalsize{$^{7}$Beijing National Laboratory for Condensed Matter Physics, Institute of Physics,}\\ \normalsize{Chinese Academy of Sciences, Beijing 100190, China}\\
\normalsize{$^{8}$School of Physical Sciences, University of Chinese }\\
\normalsize{Academy of Sciences, Beijing 100190, China}\\
\\
\normalsize{$^\ast$Email: alex.eaton@phy.cam.ac.uk}
}

\date{\today}

\begin{document}

\baselineskip24pt

\maketitle

\clearpage
\begin{sciabstract}

If a computer could be assembled from superconducting components, the energy efficiency would far surpass that of conventional electronics. Historic research efforts towards this goal yielded pivotal breakthroughs in the development and discovery of scanning tunnelling microscopy~\cite{STM82PhysRevLett.49.57} and high temperature superconductivity~\cite{bedmul86}. Although recent strides have been taken in advancing superconducting diode~\cite{SCDiode_Nature2020,SDE_review2023} and switching~\cite{matsuki2025realisation} technologies, harnessing read/writeable memory functionality in superconducting platforms has remained challenging. Here we show that bulk single crystal specimens of the triplet superconductor candidate uranium ditelluride (UTe$_2$)~\cite{Ran2019Science,Aoki_UTe2review2022,lewin2023review} possess such properties. Upon applying a magnetic field to access an intermediate regime straddling two distinct superconducting phases~\cite{Rosuel23,Aoki_Hard}, we find that direct current pulses can push the material in and out of a metastable state possessing an enhanced critical current $J_c$. This switching is controllable by the strength and duration of the stimuli, with the system `remembering' whether it is in the high or low $J_c$ state for extended periods. We interpret this to be due to competition between two distinct vortex species, which can be perturbatively pushed into a non-equilibrium high-disorder configuration with stronger pinning forces and thus higher $J_c$. Rather than requiring proximate magnetic or semiconducting interfaces~\cite{baek2014hybrid,Fermin2022,gunkel2025field,cheng2025current}, this memory functionality appears to be an intrinsic property of UTe$_2$ rooted in the superconducting order itself. Our findings underscore the rich complexity of quantum vortex matter, and demonstrate the viability of engineering a new class of superconducting memory elements with ultralow-power switching.

\end{sciabstract}

\clearpage



\noindent
Replacing resistive computational elements with non-dissipative superconducting analogues could help transform the sustainability of large-scale computing architectures -- and by extension open new horizons for their potential utility and scalability~\cite{holmes2013energy}. However, while the past decade has seen considerable progress towards engineering commercially-scalable superconducting power cable and high-field magnet technologies~\cite{yao2021superconducting,chen2022energy}, by contrast the technological exploitation of superconducting circuitry remains in its infancy. To this end, a key recent advance has been made by the experimental realisation of a superconducting diode~\cite{SCDiode_Nature2020}, which could form the backbone of superconductive logical processors~\cite{shaffer2025theoriessuperconductingdiodeeffects}. However, an experimental demonstration of electrically-controllable memory functionality, exclusively using superconducting elements, has remained elusive.

Efforts to replicate the utility of traditional magnetic storage media using superconductive components have often focussed on constructing ferromagnet-superconductor heterostructures~\cite{baek2014hybrid,Fermin2022,gunkel2025field,cheng2025current}. Such architectures can exploit spin-valve and Josephson junction functionalities, facilitating digital information encoding through physical properties such as the value of the critical current.~\cite{Birge:2016} However, magnetism is inherently antagonistic to (spin-singlet) superconductivity, and energetic dissipation in the resistive magnetic element may offset any efficiency gain made by the superconducting component. Moreover, the switching of magnetic states typically requires either the application of an external magnetic field or the generation of spin-polarised currents, which are energy-intensive and generally slower than purely electronic switching.

Beyond hybrid heterostructures, significant efforts have been directed toward vortex-trapping-based memories~\cite{chen2020miniaturization,golod2022demonstration,golod2023word}. These systems can encode information by capturing quantized magnetic flux within etched pinning sites. While such architectures avoid the dissipation associated with magnetic layers, they typically require complex lithographic engineering, such as the fabrication of sub-micron loops, and often rely on external magnetic field modulations or high drive currents for state switching. Furthermore, while several superconducting systems might themselves exhibit memory-like hysteretic behaviour due to the trapping of magnetic flux~\cite{YBCOhysteresis1989,blatterRevModPhys.66.1125,semenok2024superconducting}, these effects are generally extrinsic in nature, requiring specific geometric constraints and modulation via an applied magnetic field. By contrast, it would be highly desirable to realise an intrinsic superconducting memory effect that can be controlled solely by electrical stimuli. 


\section*{Multiphase superconductivity}

Here we study electrical transport properties of the spin-triplet superconductor candidate UTe$_2$~\cite{Ran2019Science,Aoki_UTe2review2022}. This material is a multiphase superconductor, which has numerous distinct superconducting states. These superconducting phases can be tuned between by varying the pressure alongside the strength and orientation of magnetic field~\cite{lewin2023review,Ranfieldboostednatphys2019,Braithwaite2019,aoki2020multiple,Thomas2020,Rosuel23,Kinjo_SciAdv23,tony2024enhanced,helm2024,tony2025brief,VasinaPRL25,qcl,lewin2025halo,acmsTonyPRL25}. The most easily accessible transition between two superconducting states is found at ambient pressure for a magnetic field $B$ aligned along the hard  $\hat{b}$-axis. Bulk-sensitive specific heat measurements~\cite{Rosuel23} have discerned a thermodynamic phase boundary between two superconducting states at around $B$~=~20~T. These superconductive phases are typically referred to as SC1 (the low-$B$ state) and SC2 (at higher $B$, see Fig.~\ref{fig:illustrate}a)~\cite{tony2024enhanced}. There is a substantial, growing body of evidence that SC1 and SC2 possess distinct order parameters~\cite{lewin2023review,Rosuel23,tokunaga2023longitudinal,Aoki_Hard,VasinaPRL25,acmsTonyPRL25}.

Recent measurements of the surface properties of the SC1 state at low $B$, made by scanning-tunnelling microscopy (STM) and scanning-SQUID, have revealed several anomalous vortex features. These include a mirror-asymmetric profile of the vortex cores~\cite{yang2025spectroscopic}, which form in doublets~\cite{sharma2025observation,yin2025yinyangvortexute2011}, with a tendency to organise into extended stripes of vortices~\cite{wang2025stripes}. These vortex structures form for $B>B_{c1} \approx$~4~mT~\cite{Ishihara_PhysRevResearch.5.L022002}, and so far have been found to persist up to at least 8~T~\cite{sharma2025observation}.

The vortex behaviour at higher $B$ is also unusual. Current-dependent magnetotransport measurements have discerned a region of anomalously low critical current density $J_c$ between the SC1 and SC2 states at around 15-20~T\cite{Tokiwa-vortexPRB23}. Alongside signatures in the magnetic susceptibility, these observations have been taken to indicate a region of coexistence between the SC1 and SC2 states~\cite{Aoki_Hard} that hereafter we refer to as the SC1.5 regime. Given the strong evidence that both SC1 and SC2 are odd-parity, (pseudo)spin-triplet in character~\cite{Aoki_UTe2review2022,lewin2023review}, this coexistence or mixed SC1.5 region might exhibit exotic phenomena such as fractional vortices~\cite{IvanovPhysRevLett2001} or a nonsingular vortex state~\cite{Tokiwa-vortexPRB23,Sig_Ueda91RevModPhys.63.239}.

We investigated the current--voltage ($J$--$V$) characteristics of UTe$_2$ in this region of phase space where SC1 and SC2 show signs of coexistence. When the material is entirely in either the SC1 or SC2 state, we observe forms of $V(J)$ that are typical for a type-II superconductor in a magnetic field. By contrast, in the intermediate SC1.5 region we find that abrupt perturbative modulations of $J$ lead to unusual hysteretic features in the form of $V(J)$, increasing the value of $J_c$. The system then stays in this new hysteretic high-$J_c$ state upon subsequent measurements, only returning to the original $V(J)$ profile after the application of a sufficiently large perturbation that acts to reset the system. This hysteretic tuning appears robustly long-lived (persisting over a timescale of at least several hours), constituting a novel electrically-controllable superconducting memory effect.

\section*{Superconducting memory effect in UTe$_2$}

\begin{figure}[h!]
\vspace{-1cm}
\begin{center}
\includegraphics[width=\linewidth]{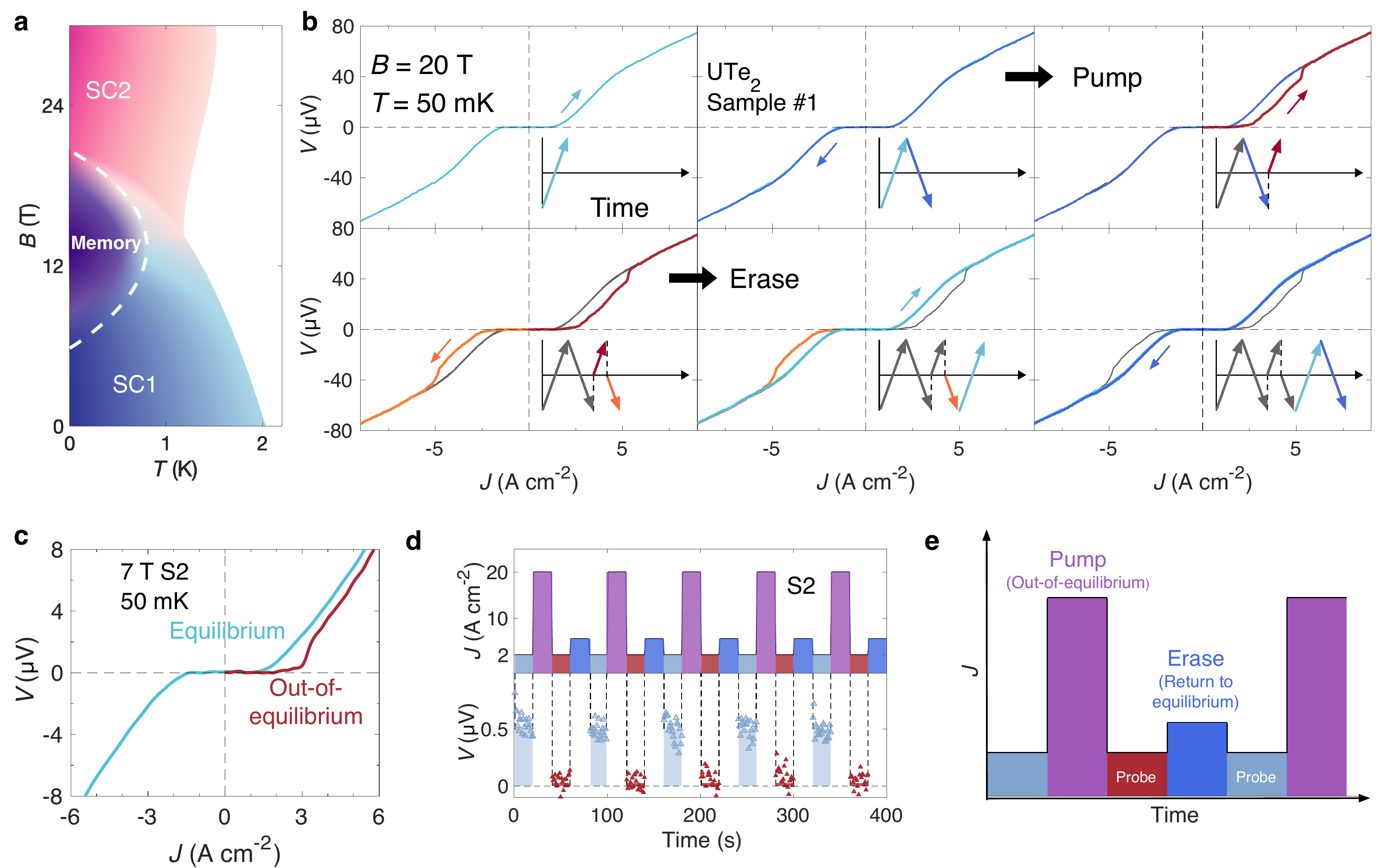}
\end{center}
\caption{\textbf{Electrical switching of an intrinsic superconducting memory effect. a,} The low temperature phase diagram of UTe$_2$ for magnetic field $B$ applied along the hard crystallographic $\hat{b}$ direction~\cite{tony2024enhanced}. The region of phase space where memory effects are manifested is coloured purple, lying between two distinct superconducting phases (SC1 and SC2). \textbf{b,} The measured voltage $V$ across a UTe$_2$ single crystal as a function of dc current density $J$. Panels are arranged chronologically from left to right, with the insets (and colour coding) showing how $J$ was modulated as a function of time. When $J$ is ramped continuously, the profile of $V(J)$ is in its equilibrium state and no hysteresis is observed. By contrast, after a discontinuous change in $J$ from a large magnitude abruptly to zero (labelled `pump'), hysteresis in $V(J)$ is observed, plotted in red and orange. Smoothly sweeping the current from a large negative value to zero (`erase') then resets the system back to the original $V(J)$ profile with lower $J_c$. All data were acquired at 50~mK with $B \parallel \hat{b}$. \textbf{c,} Data collected on sample S2 at 7~T and 50~mK. Here we plot just two traces, to highlight the hysteretic out-of-equilibrium form of $V(J)$ in the memory state. \textbf{d,} The lower panel shows $V$ measured as a function of time for modulations of $J$ depicted in the upper panel, cycling through a sequence of perturbation, measurement, and erasure (resetting) protocols, as described in \textbf{e}. By sitting at this point in the $J$--$V$ curve, successively switching in and out of the memory state yields a voltage versus time profile akin to supercurrent rectification by the superconducting diode effect~\cite{SCDiode_Nature2020}.}
\label{fig:illustrate}
\end{figure}

We plot a chronological sequence of $J$--$V$ curves in Fig.~\ref{fig:illustrate}b. Throughout this study all excitations were applied as direct currents (see \textit{Methods}), unless otherwise specified. With UTe$_2$ in its intermediate SC1.5 state at low temperatures, when the current is swept smoothly $V(J)$ is a single-valued function, with no hysteresis observed for increasing or decreasing current ramps. By contrast, after suddenly decreasing $J$ from a high value down to zero, a larger value of $J_c$ is then observed upon smoothly sweeping $J$ back up again (red curve, top right panel of Fig.~\ref{fig:illustrate}b). This hysteretic new form of $V(J)$ is also manifested for opposite polarity current ramps (orange curve, subsequent panel). Then, upon smoothly sweeping down from a large magnitude of current back to zero, the initial state is recovered, with the original form of $V(J)$ retraced for subsequent rising and falling current sweeps (blue curves). We identify this hysteretic tuning of $J_c$ as a current-controlled memory effect, and refer to the hysteretically higher $J_c$ value as the characterisation of the system being in the `memory state'.

In Fig.~\ref{fig:illustrate}c we show data measured on a second sample (labelled S2). For clarity, here we show just the non-hysteretic $V(J)$ curve (in blue, labelled equilibrium) and the post-perturbation memory state curve (in red, labelled out-of-equilibrium). Upon successive measurements, we find that the hysteretic form of $V(J)$ in the memory state persists over a timescale of at least several hours. Therefore, although the memory state is away from the equilibrium state of the system, it does not appear to be transiently metastable in character, giving promise for future development of non-volatile memory functionalities.

The tunability of the memory state enables us to perform pump/probe operations to switch between different states (Fig.~\ref{fig:illustrate}d,e). For example, if one chooses a $J$ value that lies at the bottom of the hysteresis loop, then perturbing the system to switch from the equilibrium (non-memory) to the out-of-equilibrium (memory) state changes $V$ between nonzero and zero values by raising $J_c$ (blue and red data points, Fig.~\ref{fig:illustrate}d). These high and low states could for instance be labelled 1 and 0 for computational memory purposes.

\begin{figure}[h!]
\vspace{-0cm}
\begin{center}
\includegraphics[width=\linewidth]{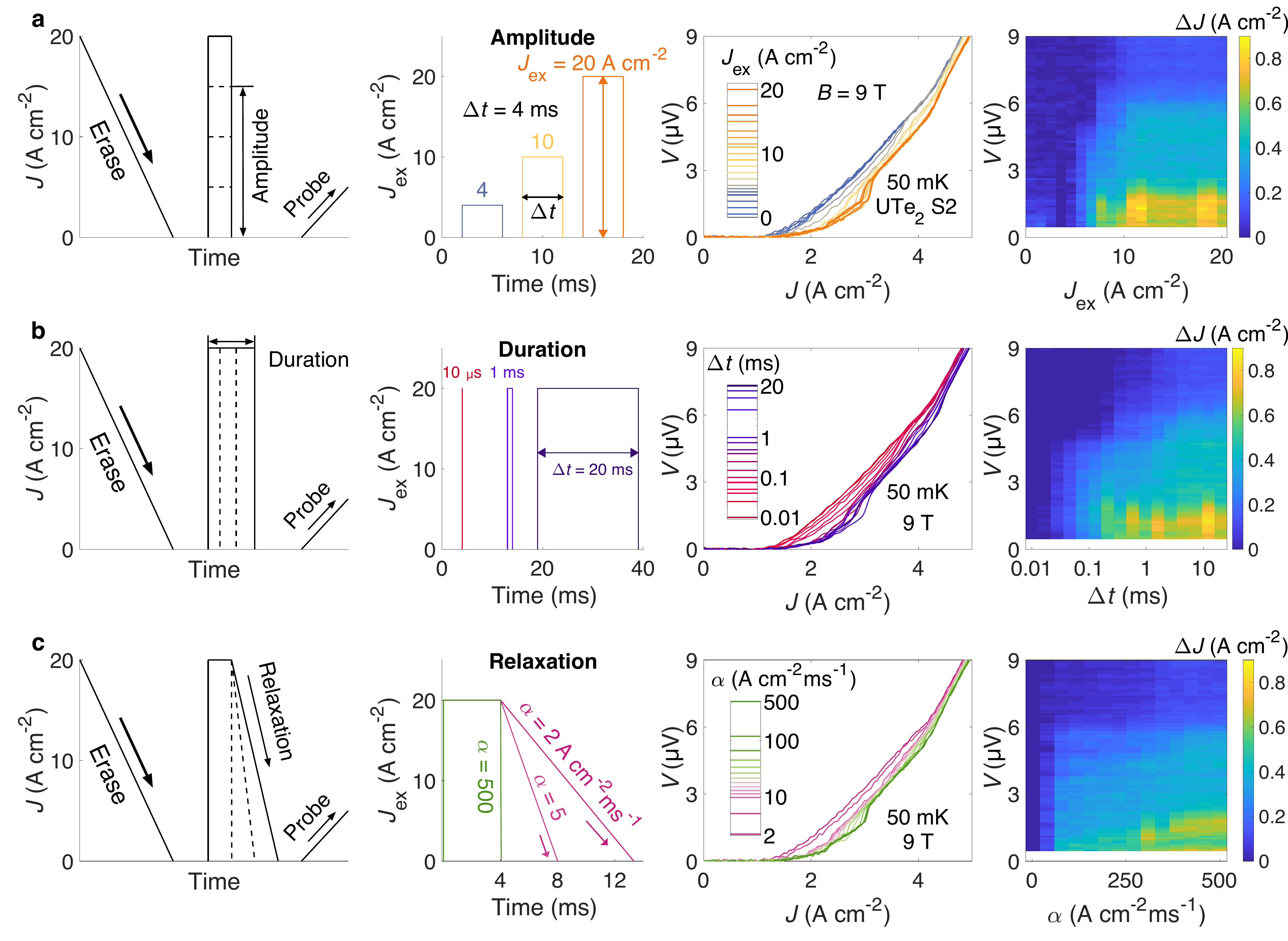}
\end{center}
\caption{\textbf{Three tuning parameters to control the memory effect.} Hysteretic response profiles for modulations of an applied excitation's \textbf{a} amplitude, \textbf{b} duration and \textbf{c} relaxation rate. In each case, the greater the perturbation to the system -- be it a larger amplitude $J_{\text{ex}}$, longer duration $\Delta t$ or more rapid relaxation rate $\alpha$ -- the greater the resulting hysteresis loop in $V(J)$, up to some saturation value (see Extended Data Fig. 1 for further analysis). For the heatmaps, $\Delta J$ is calculated as the change in $J$ at fixed values of $V$. The regions of highest $\Delta J$, coloured yellow, correspond to the strongest memory effect.}
\label{fig:tuning}
\end{figure}

\begin{figure}[h!]
\vspace{-0cm}
\begin{center}
\includegraphics[width=1\linewidth]{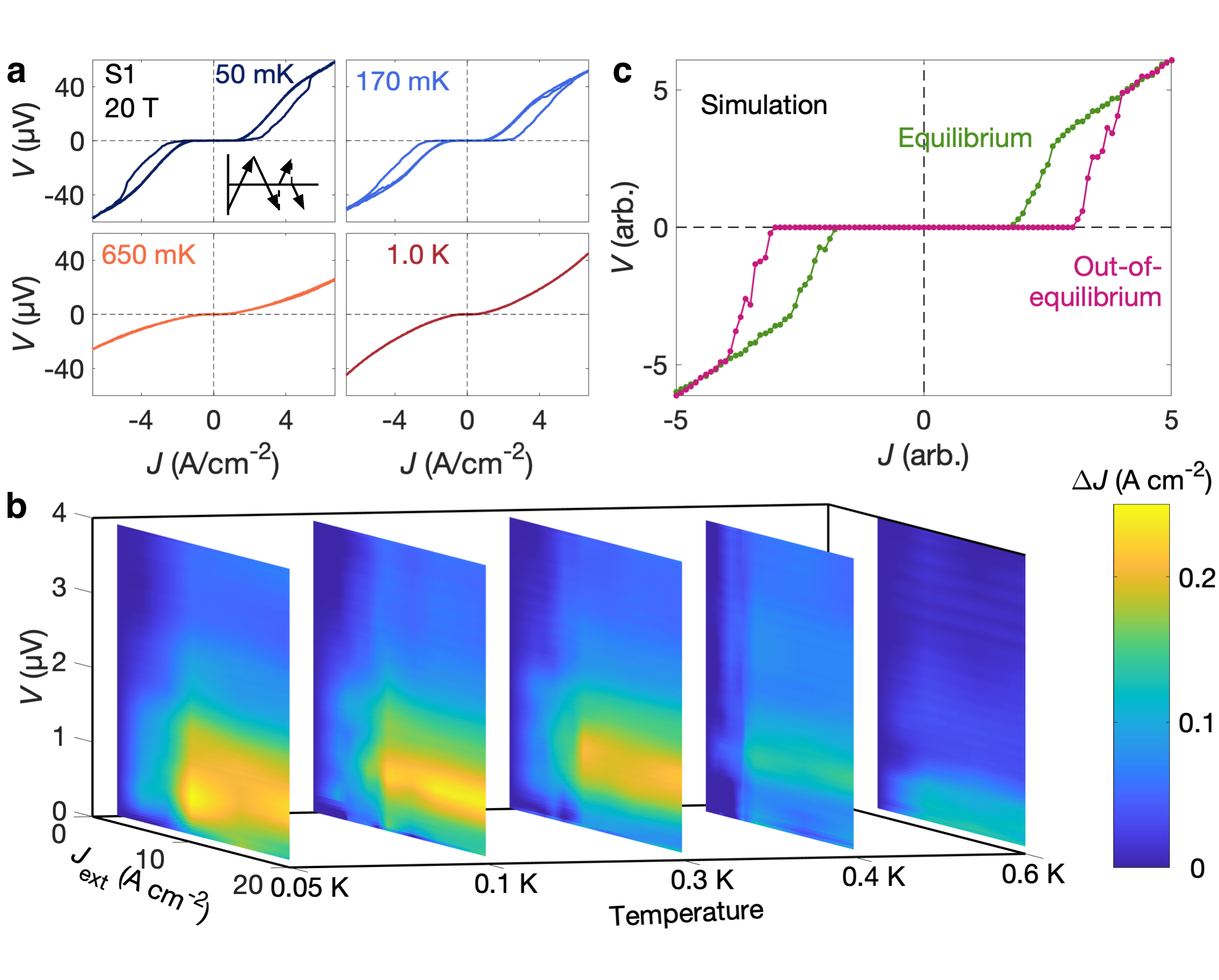}
\end{center}
\caption{\textbf{Temperature dependence and modelling of the memory effect. a,} $V(J)$ at incremental temperatures as indicated. The inset shows the sequence of current modulations. While large hysteresis loops are recorded at low temperature, these have closed by 650~mK. At 1.0~K, the curvature of $V(J)$ is markedly different compared to the equilibrium curve at 50~mK (see also Fig. S7). \textbf{b,} Heatmaps of $\Delta J$ at incremental temperatures for modulations of the excitation amplitude $J_{\text{ex}}$ up to 20~A~cm$^{-2}$ measured on sample S2 at 14~T. Again, yellow colouring indicates the strongest memory effect, which sharply diminishes at elevated temperatures. \textbf{c,} Simulated $J-V$ curves for the two types of current pulses, which correspond to the equilibrium and non-equilibrium switching of current, see \textit{Methods}. The memory state is when the system is out-of-equilibrium. The calculated curves are in good qualitative agreement with the measured low temperature data.}
\label{fig:temps}
\end{figure}

We find that the memory effect can be controlled by tuning the system in three different ways: by modulating the amplitude, duration and relaxation rate of the perturbative excitation current $J_{\text{ex}}$ (Figure~\ref{fig:tuning}). For each of these tuning parameters, we find that larger hysteresis loops are induced for stronger perturbations, up to some saturated maximal values (see Extended Data Fig. 1). Furthermore, each of these tuning axes yields a similar evolution in temperature of the memory effect. Figure~\ref{fig:temps} depicts the temperature $T$ dependence over the interval 50~mK $\leq T \leq$ 1.0~K for amplitude-tuning. At low temperatures below 0.4~K bright yellow regions in Fig.~\ref{fig:temps}b identify a strong memory effect characterised by the observation of a large $\Delta J$ for high $J_{\text{ex}}$. However, at elevated temperatures this quickly diminishes, with hysteresis loops closing for $T >$ 0.6~K. Similar behaviour is also observed under duration- and relaxation-tuning, as shown in Extended Data Fig.~5.

Such a low temperature scale of the memory effect in UTe$_2$ is somewhat surprising, given that the superconducting $T_c$~=~2.1~K in zero field, and is still $>$~1~K in this field range. This points to some intrinsic low energy phenomenon, distinct from the superconductivity itself, being responsible for the observed hysteretic $V(J)$ phenomena. It appears analogous to the superconducting diode effect, for which extrinsic vortex-driven effects can be enhanced near $T_c$ but diminish at lower $T$. By contrast, for intrinsic mechanisms (in which the superconducting state itself spontaneously breaks inversion and time-reversal symmetries) the diode effect becomes enhanced in the low $T$ limit~\cite{Yanase-intrinsicPRL22}, similar to our observations here of the superconducting memory effect in UTe$_2$.

\section*{Mapping the memory effect}

\begin{figure}[h!]
\vspace{-1cm}
\begin{center}
\includegraphics[width=1\linewidth]{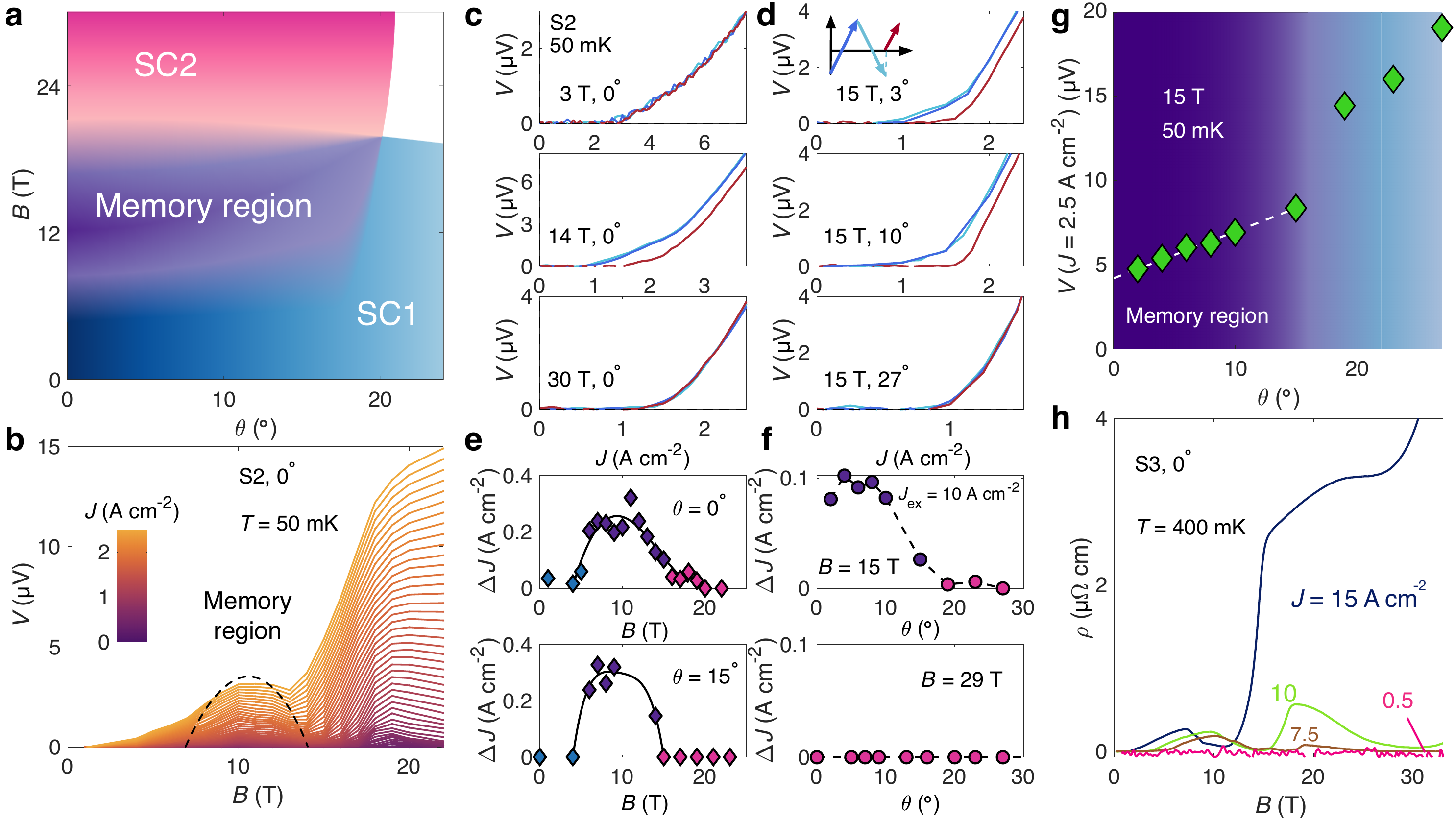}
\end{center}
\caption{\textbf{Mapping the memory region between SC1 and SC2. a,} Schematic phase landscape of UTe$_2$ for rotations of $B$ by an angle $\theta$ from $\hat{b}$ towards $\hat{c}$. The memory region (coloured purple) -- identified by nonzero $\Delta J$ at low $T$ -- is located at the intersection of the SC1 and SC2 domains. We refer to this as SC1.5. \textbf{b,} The equilibrium profile of $V(B)$ at 50~mK for successive $J$ values, as indicated by the colour scale. An anomalous non-monotonic kink in $V(B)$ at high $J$ is observed at the boundary of the memory region. \textbf{c,} Sequential modulations of $J$ at different fields for $B \parallel \hat{b}$ (0$\degree$) and \textbf{d,} modulations of $J$ at different magnetic field tilt angles $\theta$ at $B =$~15~T. The inset defines the measurement protocol. \textbf{e,} Evolution of $\Delta J$ as a function of $B$ at 0$\degree$ and 15$\degree$ and \textbf{f,} as a function of $\theta$ at 15~T and 29~T. \textbf{g,} Equilibrium $V(\theta)$ for $J$~=~2.5~Acm$^{-2}$, $B=$~15~T, $T=$~50~mK measured on sample S2. The transition from the memory region into SC1 is marked by a sudden jump in $V$ at around $\theta =$~20$\degree$, indicating a change in the vortex properties. This angle is where, in panel (f), $\Delta J$ reaches zero within the resolution of the measurement.  \textbf{h,} Magnetic field sweeps of the effective resistivity $\rho$ for incremental $J$ values as indicated. The data in this panel were acquired by low frequency ac measurements, whereas all other data in this article are from dc measurements (see \textit{Methods}). While zero resistance is observed over all $B$ in the low $J$ limit, at higher $J$ a complex profile of flux-flow behaviour is exhibited.}
\label{fig:map}
\end{figure}

To reveal the full domain of phase space in which memory effects are present, we performed a detailed mapping of the hysteretic $V(J)$ characteristics of UTe$_2$ as a function of magnetic field strength and rotation angle up to 30~T at 50~mK, which is summarised in Figure~\ref{fig:map}. At low $B$ the form of $V(J)$ is single-valued and does not exhibit hysteresis under perturbative amplitude-, timescale- or relaxation rate-tuning. As $B$ is increased, hysteretic signatures of the memory effect start to appear at around 7~T, which persist up to higher $B$ until the phase boundary out of SC1 into SC2 is crossed. The exact value of $B$ at which this occurs varies slightly from sample to sample depending on crystalline quality~\cite{Rosuel23,tony2024enhanced} -- for sample S1 this is just above 20~T, whereas for S2 it is just below 20T (see Fig.~\ref{fig:map}e and Extended Data Figs. 8 \& 9). At high $B$ around 30~T the material is then fully in the SC2 phase, and the memory effect is no longer exhibited (Fig~\ref{fig:map}c).

Tilting $B$ provides insight into the character of the memory effect. The SC2 phase of UTe$_2$ only resides in a narrow range of $\theta$ close to $B \parallel \hat{b}$, which for high quality salt flux-grown crystals such as those studied here extends out to $\theta \approx 20\degree$~\cite{tony2024enhanced}. We find that the memory region closely tracks this. In Fig.~\ref{fig:map}e\&f we plot the value of $\Delta J$ observed for a perturbative $J_{\text{ex}}$ for various $B$ and $\theta$ (see Extended Data Fig. 7 for raw $V(J)$ curves). At 0$\degree$ no change in $\Delta J$ is observed at low $B$, which quickly rises for $B >$ 7~T and peaks just above 0.2~A cm$^{-2}$ at around 10~T. A very similar profile is exhibited at $\theta = 15\degree$, indicating that the strength of the memory effect is not strongly peaked at the $\hat{b}$ axis but is similar in magnitude throughout the $B-\theta$ parameter space directly below SC2. Then, upon rotating beyond the range of SC2, $\Delta J$ abruptly diminishes back to zero (Fig.~\ref{fig:map}f). This truncation of the memory region is also reflected in the variation of $V$ measured in equilibrium for a fixed $J$ (Fig.~\ref{fig:map}g), which suddenly jumps upon departing the memory region into exclusively SC1 close to $\theta = 20\degree$. The close connection between the angular range in which SC2 sits above SC1, and the  observed domain of phase space that hosts the memory effect, underscores that the interplay of these two distinct superconducting states is imperative in generating the observed hysteretic $V(J)$ phenomena.


\section*{\label{sec:results}Modelling the memory effect}

Now we turn to considering the likely microscopic origin of the memory effect in UTe$_2$. First, given that we are applying significant electrical current pulses at dilution fridge temperatures, we should consider any extrinsic heating effects that might be present. In the \textit{Supplementary Information} we plot the measured temperature as a function of time during the current pulse measurement protocols, and give further discussion as to why we do not believe that any Joule heating effects significantly affect our measurements. One strong argument against any Joule heating artifacts comes from the sensitivity of the memory effect to the relaxation rate of the dc stimulus. In Fig.~\ref{fig:tuning} we showed that stronger memory effect correlates with larger amplitude and duration tuning. For both of these, the larger stimulus correlates with more energy deposited into the system (through lead/contact resistance). However, the memory effect is also strongly correlated with faster relaxation rates $\alpha$, which \textit{anticorrelates} with the total energy transferred into the system. If the memory effect were due to some relatively straightforward Bean-Livingston barrier or Bean model pinning effect process~\cite{Tinkham04}, this should be insensitive to how the current is turned off, depending only on the total energy imparted. The acute sensitivity of the memory effect to the turn-off speed instead implies some dynamic re-organisation or annealing of vortex domains, likely due to some interesting intrinsic physical property (or properties) of UTe$_2$.

Further arguments against heating effects stem from the fact this phenomenon is only observed in the narrow regime of $B-\theta$ space directly underneath SC2, whereas any extrinsic heating phenomena should likely be present throughout SC1 and SC2.  The memory effect is most pronounced at the lowest temperatures, and quickly diminishes upon warming (Fig.~\ref{fig:temps}), providing perhaps the strongest argument against any heating-induced artifacts. Instead, this strongly indicates that the observed hysteretic $V(J)$ behaviour is due to some intrinsic low-energy phenomenon, with an energy scale significantly lower than $T_c$.

We propose that the memory effect may be due to competition between two different vortex species -- one native to SC1, and the other from SC2. In the equilibrium state, the vortices arrange themselves in a highly ordered configuration to minimise entropy and energetic costs. Upon perturbing the system, current densities above $J_c$ are applied, inducing a flux-flow regime in which both types of vortex are dissipatively travelling through the sample. If the perturbation is sufficiently sharp -- for example, if the relaxation rate $\alpha$ is very fast -- then when $J$ drops back below $J_c$ vortex flow suddenly halts and the vortices are abruptly locked into their new positions. One can imagine a glassy vortex state has now been quenched in the material, which possesses much higher disorder than the original equilibrium vortex lattice. Higher disorder implies stronger pinning forces between vortices, and hence the higher $J_c$ values that characterise the memory effect. The system can then be reset by smoothly raising $J$ back above $J_c$ before gradually ramping back down to zero current. This could be thought of as annealing the system, thereby returning to the equilibrium low-$J_c$ state.

To explore this idea of quenching and annealing vortex configurations, we constructed a resistively-shunted-junction (RSJ) model for the UTe$_2$ memory effect. Such modelling is often used to describe transport \cite{BuzdinPRB2007, AmbegaokarHalperin1969} and tunnelling \cite{Devoret85, Devoret87} in Josephson junctions, but also captures the phenomenology of current dynamics in junction-free samples\cite{Tinkham04, Shapiro92}. We assumed that the total current \(J=J_S+J_N+J_n\) may be decomposed into a supercurrent \(J_S(\phi)\), normal current \(J_N=\gamma V\) with damping \(\gamma\sim1/R\), and a noise current \(J_n\) that can account for vortex flow when vortices become unpinned (see \textit{Methods} for full calculation details). The supercurrent is a function of the superconducting phase \(\phi\) applied across the superconductor \(J_S(\phi)=U'(\phi)\), where \(U(\phi)\) is the energy of the superconductor and the prime denotes differentiation with respect to \(\phi\) (see Extended Data Fig. 2), satisfying the Josephson relation \(\dot{\phi}=2eV/\hbar\). 

The RSJ equations are analogous to the equation of a pendulum subject to a torque \(\tau\sim J\), with \(\phi\) corresponding to the deflection angle of the pendulum, \(U(\phi)\) to the gravitational potential, \(\gamma\) to the damping, and \(\dot{\phi}\propto V\) to the angular velocity of the pendulum. The nonzero voltage state then corresponds to a steady state in which the pendulum completes full revolutions. Similarly, this treatment is also analogous to the description of a particle with position variable \(\phi\) in a tilted washboard potential \(U(\phi)-J\phi\).


Within this minimal model, the memory effect may be understood if \(U(\phi)\), which has a global minimum at \(\phi=0\) (implying no supercurrent in the absence of an applied phase gradient), has a second local minimum at \(\phi\neq0\), with \(\Delta(r)\sim e^{i\phi r/L}\) where \(L\) is the length of the effective unit cell of that state (see Extended Data Fig.~2). As inversion symmetry appears unbroken in the metastable state, this second minimum has to occur at \(\phi=\pi\), corresponding to an inhomogeneous SC state in the SC1.5 region. 
This might consist of a grid of domain walls, in which the walls separate local regions within which the material is either in the SC1 or SC2 state. (More generally, this could also be interpreted as mesoscopic regions with different vortex species\cite{MingarelliPhysRevLett19} whereby \(L\) would correspond roughly to the average diameter of the domains.) Regardless of the precise nature of the vortices' configuration and any possible domain structure, the coexistence of multiple order parameters is crucial for realising a non-trivial minimum of \(U(\phi)\). This indicates why the memory effect is only seen in a narrow region of $B-\theta$ space (Fig.~\ref{fig:map}) where SC1 and SC2 coexist. Furthermore, larger damping would be consistent with the scenario of a higher concentration of domain walls in the memory state, corresponding to a lower resistance \(R\sim1/\gamma\) in the vortex flow state \cite{Kim65, BlatterLarkinRMP94, Reichhardt25}. Therefore, due to enhanced pinning at domain boundaries, this naturally accounts for the higher value of $J_c$ in the memory state.


We confirm that this form of \(U(\phi)\) indeed captures the phenomenology of the memory effect by direct simulation of the dynamical RSJ equations (Fig.~\ref{fig:temps}c and Extended Data Fig. 2). Under a pump/probe (short current \(J(t)\) pulse) protocol, \(\phi(t)\) is found to land in the \(\phi=\pi\) minimum after the pump regardless of its initial condition, while under an erasure (slow current decrease) protocol \(\phi(t)\) tends to reach the global minimum. Microscopically, the latter sequence can be thought of as an annealing process in which the vortices relax into more stable (equilibrium) positions, as proposed for example in ref. \cite{Reichhardt25}.

\section*{Outlook}

An attractive technological use case for the superconducting memory effect is as an ultralow-power alternative to solid state or hard disk drive memory functionality. Moreover, the inherent plasticity of the effect presents a promising route towards cryogenic neuromorphic processing. For the bulk mm-sized single crystal specimens investigated here, to probe (read) whether the material is or isn't in the memory state -- which could correspond to a binary 1 or 0 -- required a power consumption of $\sim$~1~nW. Scaling down even just mesoscopically to the order of one square micron in size would reduce this to $\lesssim$~1~pW -- orders of magnitude less than for nanometre-sized silicon elements~\cite{green1990intrinsic}. Furthermore, whereas cryotron-based~\cite{buck1956cryotron} superconducting devices require transitioning to the normal state, UTe$_2$ has the distinct advantage that both read and write operations can be performed while superconducting. Recent studies have demonstrated this material's ability to be carved at the micron-scale by focused ion beams~\cite{helm2024,ZhangMoll25}, making the prospect of lithographically processing UTe$_2$ wafers~\cite{yoon2026submicrometer} a realistic albeit ambitious goal.


Further work is required to assess whether similar memory phenomena could be exhibited by other multiphase superconductors such as CeRh$_2$As$_2$~\cite{khim2021CRA} or UPt$_3$~\cite{TailleferUPt3RevModPhys.74.235}. We hope that our discovery here may catalyse efforts to discover new multiphase superconductors composed of more abundant elements, which might exhibit the memory effect to higher temperatures. Although the list of known multiphase superconductors is short, notable recent additions from materials as diverse as moir\'e graphene~\cite{cao2021pauli} and high-$T_c$ nickelates~\cite{rubi2025extrememagneticfieldboostedsuperconductivity} suggest that further promising candidates await discovery.

In summary, we studied the current-voltage characteristics of the multiphase superconductor UTe$_2$. In an intermediate phase space regime bordering two superconducting phases, we observed anomalous hysteresis in the value of the critical current, dependent on the history of applied direct current ramps and pulses. We showed that modulating the amplitudes and timescales of dc electrical stimuli can selectively tune between low and high critical current states, with the material then `remembering' which state it is in. We interpret this to be due to different vortex configurations being either abruptly quenched or smoothly annealed. Our work demonstrates that, in principle, the vortex matter of an unconventional superconductor can be electrically manipulated and used to encode and process information, thereby opening up new possibilities for low energy electronic devices, cryogenic neuromorphic memory, and quantum computational hardware.

\clearpage
\normalsize{
\bibliographystyle{naturemag_noURL}
\bibliography{UTe2}
}

\clearpage
\Large
\noindent
\textbf{Methods}

\normalsize
\vspace{5mm}\noindent 
\textbf{Sample preparation}\\ Single crystal UTe$_2$ specimens were grown in a flux of molten salts by the recipe detailed in ref.~\cite{Eaton2024}. Samples were contacted by spot-welding four gold wires, each of 50~$\upmu$m diameter, for respective current and voltage leads. These were subsequently connected to copper twisted pairs, which were soldered to the wiring of the probe. Each of the three samples investigated in this study had $J$ sourced along the [100] direction, with leads affixed to the (001) surface. Residual resistivity ratio (RRR) measurements were performed using a Quantum Design Physical Properties Measurement System in Cambridge with low frequency ac excitations. The RRR values for the three samples are S1: 105; S2: 44; S3: 395. 

\normalsize
\vspace{5mm}\noindent 
\textbf{Electrical transport measurements}\\ The hysteretic $J-V$ properties of UTe$_2$ were investigated by performing direct current measurements using a Keithley 6221--2182A combined current source--multimeter system. The in-built pulse-delta sweep mode and arbitrary waveform generator features were utilised. Measurements were performed in a 30~T superconducting magnet with Oxford Instruments dilution fridge system at the Synergetic Extreme Condition User Facility (SECUF), Beijing. All measurements presented in this study were performed in that system, except those in Fig.~\ref{fig:map}h and Fig. S2 that were performed at the National High Magnetic Field Lab, Florida, in a resistive magnet utilising a $^3$He system, in which low-frequency ac measurements were obtained using a Keithley 6221 current source and a Stanford Research 860-series lock-in amplifier. For the electrical switching study at SECUF, a sequence of pulse-delta measurements with different applied currents was defined by entering the starting current value, ending current value, and the number of steps for the sweep. To control the amplitude, duration, and relaxation rate, a varying shape excitation pulse was defined by 100 points and generated with the arbitrary waveform feature of the source--multimeter system. Control of the instrument was realised by custom-developed Python code with the pymeasure package~\cite{Pymeasure_source}, which is included as a supplement to this Article. The reset protocol prior to performing a perturbative excitation was achieved by sweeping the current from 40~mA to 0~mA smoothly with a 2~mA step over a 20 second time interval with the pulse-delta sweep mode. The cycling sequence in Fig.~\ref{fig:illustrate}d was performed by the pulse-delta mode. The time sequence is defined by entering the current amplitude and number of steps.

\normalsize
\vspace{5mm}\noindent 
\textbf{Modelling of the memory effect}\\ As introduced in the main text, we employed a modified resistively-shunted-junction (RSJ)-analogue model, similar to ref.~\cite{BuzdinPRB2007}, which describes a Josephson junction by considering a potential with two minima, see Extended Data Fig. 2. We neglected capacitive effects as no hysteresis is seen in the absence of a pump (corresponding to an overdamped Josephson junction limit in the RCSJ model). The potential that describes the model is given by 
\begin{equation}
    U (\phi) = \left( 1 - \cos \phi + g\frac{1 - \cos 2 \phi}{2} \right) \left( 1 - 0.6 \left( \frac{1 - \cos \phi}{2} \right)^{\beta} \right).
    \label{eqn:U}
\end{equation}
This potential has two minima: $\phi=0, \pi$. Such choice of phases is consistent with the inversion symmetry as in the RSJ-analogue model the phase $\phi$ corresponds to the Cooper pair momentum in the time-dependent Ginzburg-Landau description. The global minimum at $\pi=0$ is the true ground state of the system; the minimum at $\phi=\pi$ corresponds to a long-lived metastable ``glassy'' state, which represents the memory state of UTe$_2$. Parameter $\beta$ controls the critical current of the local minimum at \(\phi=\pi\). In our calculations we chose $\beta = 40$, such that the minimum at \(\phi=\pi\) has a higher critical current than the global minimum at \(\phi=0\).

The memory effect is a dynamical effect involving the time evolution of $\phi (t)$. This time dependence is determined by the equation
\begin{equation}
     J = J_S+J_N+J_n=U'(\phi)+\gamma(\phi)\dot{\phi}+J_n
    \label{PDE}
\end{equation}
where \(\dot{\phi}=d\phi/dt\), \(J_S=U'=dU/d\phi\) and \(\gamma(\phi)=\gamma_0+ \gamma_\pi \left( \frac{ 1 - \cos \phi }{2}  \right)^{\delta}\). The noise current \(J_n\) satisfies \(\langle J_n(t)J_n(t')\rangle\sim\delta(t-t')\) (uncorrelated in time \(t\)); in principle this treatment may be derived using a time-dependent Ginzburg-Landau theory \cite{Tinkham04,Shapiro92,BuzdinPRB2007,AmbegaokarHalperin1969}. The parameters $\gamma_0$ and \(\gamma_\pi\) (both chosen to be \(1\) in simulations) control the ``viscosity'' at the two minima, which determines how likely the system is to get stuck in a given minimum after a pump. Note that larger vortex flow damping \(\gamma\) corresponds to lower velocity of the vortices, and therefore lower dissipation and resistance; in particular, vortices are pinned at infinite damping, corresponding to a zero resistance state. The parameter $\delta$ defines how localised the viscosity is and which of the two minima is ``viscous''. We chose our viscosity in such way that the glassy (metastable) minimum at $\phi = \pi$ is viscous and used $\delta = 20$ in our numerical simulations. The large damping \(\gamma(\phi)\) at the local minimum must be a function of the phase \(\phi\) peaked around \(\phi=\pi\). This makes \(\phi(t)\) tend to `stick' more easily around \(\phi=\pi\) when it evolves dynamically, despite \(\phi=0\) being energetically more favourable.

To study the dynamics of the junction we numerically solve Eq. \eqref{PDE} with initial conditions $\phi (0) = 0, \pi$ for different current driving regimes. The two types of driving we consider are: (i) a short rectangular current pulse (quench; out-of-equilibrium regime) and (ii) a slowly varying triangular current pulse (``annealing'' or ``erasure''; quasi-adiabatic evolution), mimicking the experimental setup. Within our effective description the memory effect is achieved by the switching between different minima of the effective potential: for the regime (i) starting from either of the two minima, the system ends up in the minimum corresponding to $\phi=\pi$. Conversely, for the erasure regime (ii), the system ends up in the $\phi=0$ minimum regardless of the initial $\phi$. Such switching can be seen in plots of the explicit time dependence of the phase $\phi (t)$ in Extended Data Fig. 2d,e. We interpret this non-equilibrium switching due to the quench to be associated with the formation of a long-lived transient (metastable) state\cite{ChichinadzeTransientPRB2016}. Furthermore, many of the properties of the intermediate SC1.5 regime could also be consistent with a finite-momentum paired state (\(\Delta(r)\sim \cos q r\)), which has been proposed to explain some experimental observations of UTe\(_2\). \cite{ZhangMoll25, wang2025stripes, Aoki_Hard}. This has also been theorised to realise a superconducting diode effect \cite{DanielSDE2024}.

To plot the $J-V$ curves we find the asymptotic behaviour of $\frac{d\phi(t)}{dt} \propto V$ on large timescales, so that the system reaches a (meta)stable state, and treat $\gamma$ as a proxy to the injected current $J$, see Extended Data Fig. 2. To avoid a known issue of the voltage jump at the critical current value we follow ref.\cite{AmbegaokarHalperin1969} and add noise to our simulation and average over $N$ resulting $J-V$ curves, where $N$ is the number of solutions for different noise realisations. In our work we averaged over $N=60$ noise realisations.

\Large
\vspace{12mm}\noindent
\textbf{Acknowledgements}

\normalsize
\vspace{5mm}\noindent 
\noindent
We gratefully acknowledge stimulating discussions with D. Agterberg, E. Babaev, P. Coleman, J. Durrell, A. Greer, A. Huxley, P. Littlewood, G. Lonzarich, C. Reichhardt, T. Winyard and especially A. Levchenko. This project was supported by the EPSRC of the UK through grants EP/X011992/1 and EP/R513180/1. A portion of this work was carried out at the Synergetic Extreme Condition User Facility (SECUF, \href{https://cstr.cn/31123.02.SECUF}{https://cstr.cn/31123.02.SECUF}). A portion of this work was performed at the National High Magnetic Field Laboratory, which is supported by National Science Foundation Cooperative Agreement Nos. DMR-1644779 \& DMR-2128556 and the State of Florida. We acknowledge financial support by the Czech Science Foundation (GACR), project No. 22-22322S. The work of D.S. was financially supported by the NSF Quantum
Leap Challenge Institute for Hybrid Quantum Architectures and Networks Grant No. OMA-2016136. D.V.C.
acknowledges financial support from the National High Magnetic Field Laboratory through a Dirac Fellowship, which is funded by the National Science Foundation (NSF) (Grant No. DMR-2128556) and the State of Florida, and from Washington University in St. Louis through the Edwin Thompson Jaynes Postdoctoral Fellowship. T.I.W. acknowledges support from Murray Edwards College (University of Cambridge) and the Cambridge Philosophical Society through a Henslow Fellowship. T.I.W. and A.G.E. acknowledge support from ICAM through US National Science Foundation (NSF) Grant Number 2201516 under the Accelnet program of the Office of International Science and Engineering and from QuantEmX grants from ICAM and the Gordon and Betty Moore Foundation through Grants GBMF5305 \& GBMF9616. A.G.E. acknowledges support from Sidney Sussex College (University of Cambridge).

\vspace{12mm}
\Large
\noindent
\textbf{Data availability}

\normalsize
\vspace{5mm}\noindent 
The datasets supporting the findings of this study will be uploaded to the University of Cambridge Apollo Repository prior to publication.

\vspace{12mm}
\Large
\noindent
\textbf{Competing interests statement}

\normalsize
\vspace{5mm}\noindent 
The authors declare no competing interests.

\clearpage

\begin{figure}[t!]
    \vspace{2cm}
    \begin{center}
    \includegraphics[width=.8\linewidth]{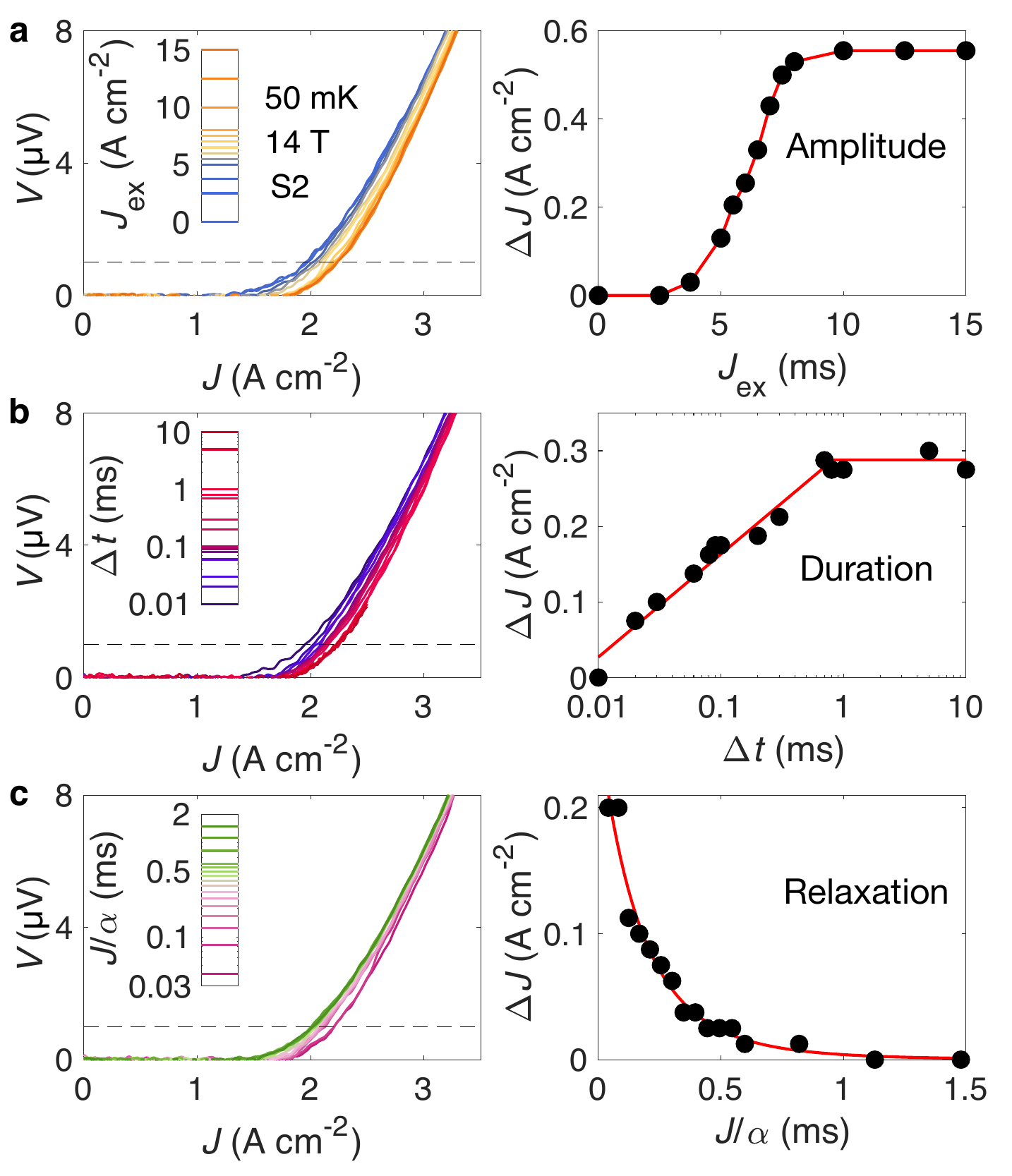} 
    \end{center}
    \vspace{-5pt}
    \caption*{\textbf{Extended Data Fig. 1 | Saturation of the memory effect. a,} Amplitude, \textbf{b,} duration and \textbf{c,} relaxation-rate tuning of the memory effect of UTe$_2$. For sufficiently large perturbative amplitudes $J_{\text{ex}}$ or durations $\Delta t$, the memory effect approaches a saturation value. $\Delta J$ is calculated here at a nominal fixed value of $V =$~1~$\upmu$V as indicated by the dashed lines. All data were acquired on sample S2 at 14~T and 50~mK.}
    \label{pic:14T_tuning}
\end{figure}


\begin{figure}[h!]
    \vspace{0cm}
    \begin{center}
    \includegraphics[width=.9\linewidth]{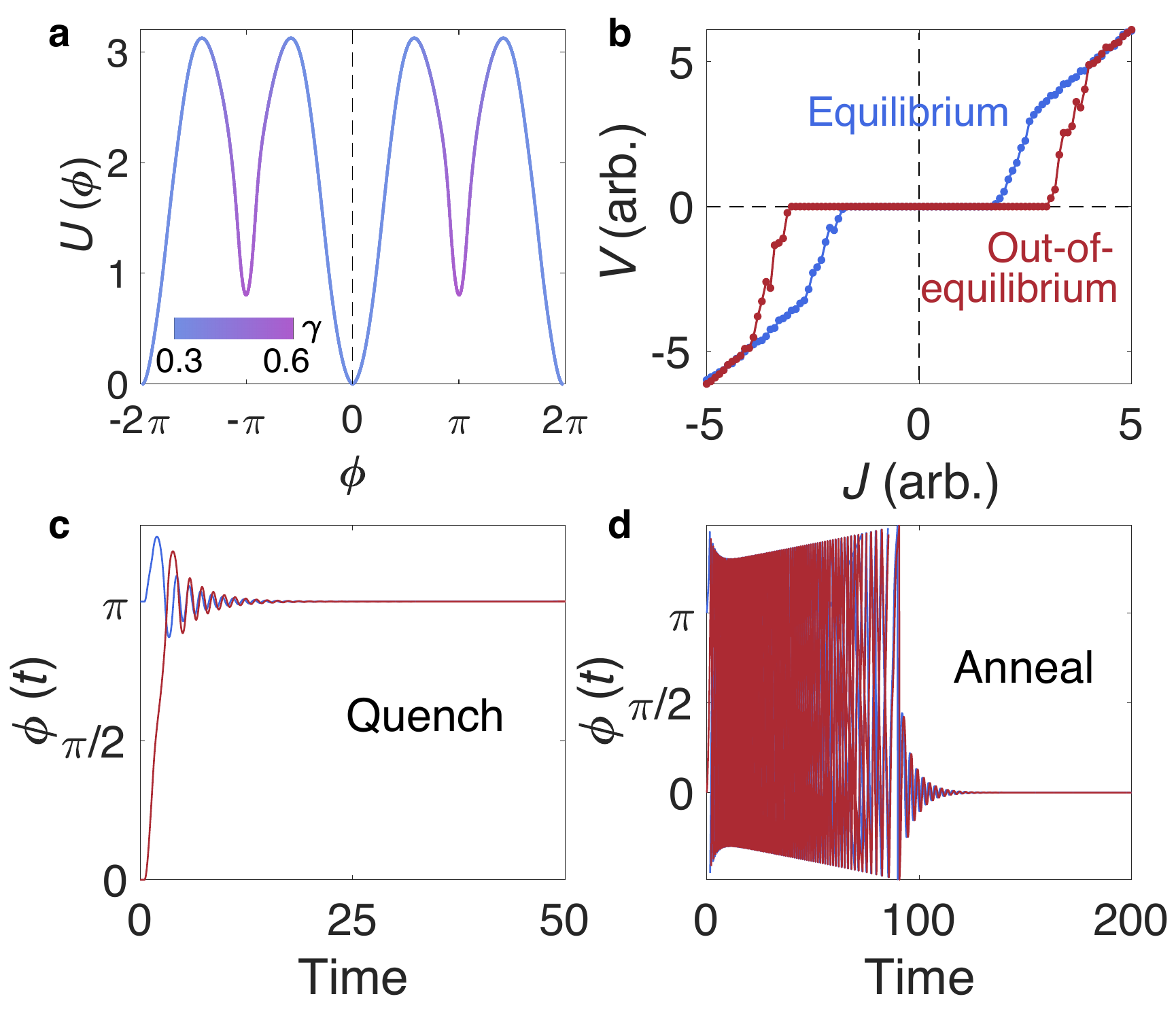} 
    \end{center}
    \vspace{-5pt}
    \caption*{\textbf{Extended Data Fig. 2 | Theoretical modelling of the memory effect. a,} Plot of the potential $U(\phi)$ from Eqn.~\ref{eqn:U}. Note the two different minima: global at $\phi=0$ and the local ``glassy'' minimum at $\phi = \pm \pi$. The colour code illustrates the value of the parameter $\gamma$ characterising ``friction''. \textbf{b,} The resulting $J-V$ characteristic for the case with random noise and averaged over $N=60$ noise realizations. \textbf{c,} Time evolution of $\phi(t)$ for the short current pulse (``quench'') and \textbf{d,} same as in the previous panel but for the erasing current (``annealing'').}
    \label{pic:Theory}
\end{figure}

\begin{figure}[t!]
    \vspace{-2cm}
    \begin{center}
    \includegraphics[width=.7\linewidth]{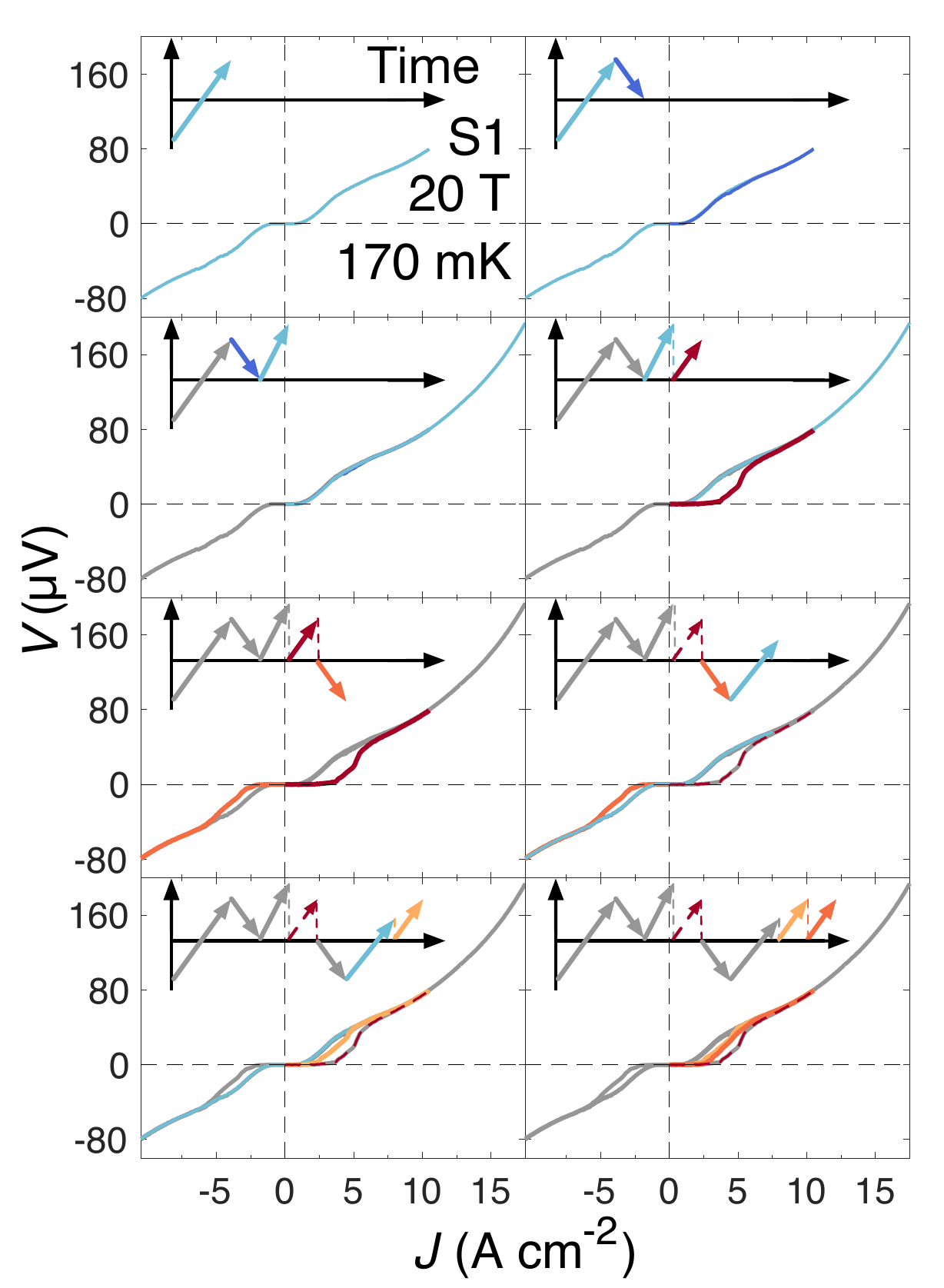} 
    \end{center}
    \vspace{-5pt}
    \caption*{\textbf{Extended Data Fig. 3 | Discrete tuning of memory hysteresis.} Controllability of the superconducting memory effect by modulation of the perturbation amplitude. Insets show the chronological sequence of how $J$ was swept. In Fig.~\ref{fig:illustrate} we always perturbed the system by discontinuously changing $J$ from the same magnitude -- by contrast, here we perform the perturbation from two different values of $J$. A larger hysteresis in $V(J)$ is observed for the larger perturbation. This shows that the magnitude of the memory effect is tunable by the size of the excitation, similar to Fig.~\ref{fig:tuning}.}
    \label{pic:14T_tuning}
\end{figure}



\begin{figure}[t!]
    \vspace{2cm}
    \begin{center}
    \includegraphics[width=.8\linewidth]{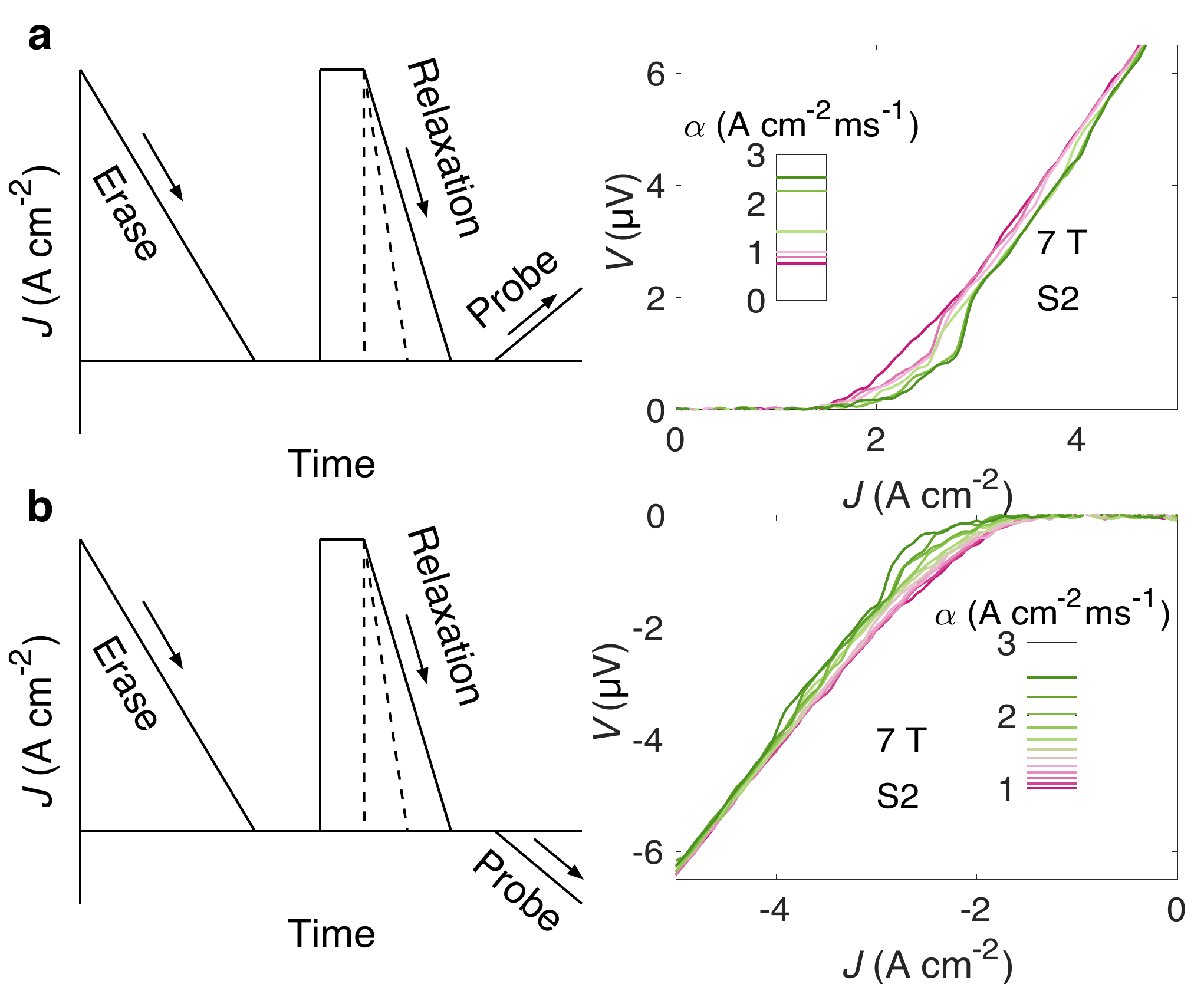} 
    \end{center}
    \vspace{-5pt}
    \caption*{\textbf{Extended Data Fig. 4 | Presence of the memory effect under both polarities of current flow. a,} Results for positive and \textbf{b,} negative polarity of $J$. The same polarity of erasure and perturbation were performed in both instances, but then measured by opposite-polarity probe measurements (as depicted in left-hand-side schematics). Consistent behaviour is observed for both orientations.}
    \label{pic:PENP}
\end{figure}

\begin{figure}[h!]
    \vspace{2cm}
    \begin{center}
    \includegraphics[width=1\linewidth]{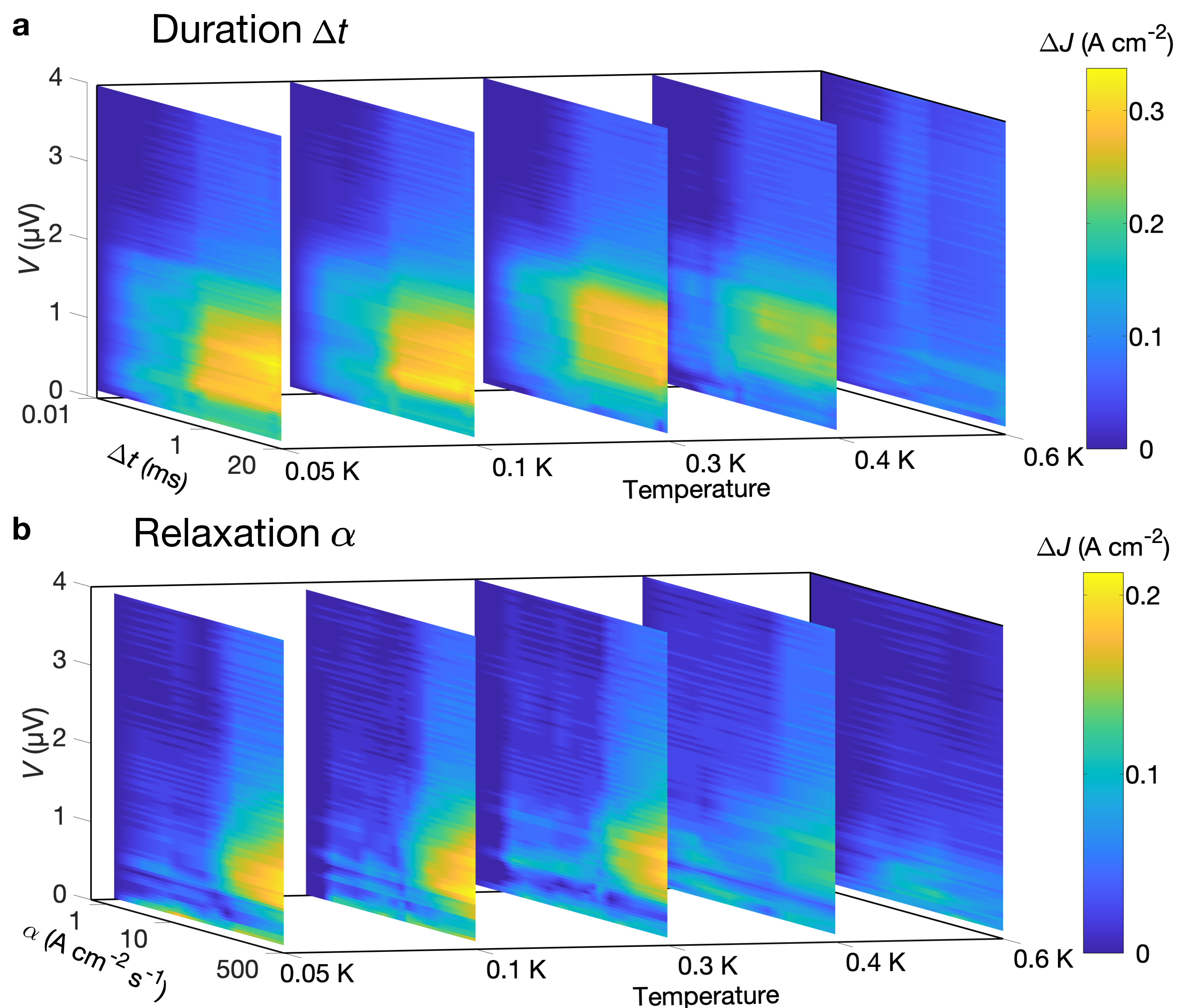} 
    \end{center}
    \vspace{-5pt}
    \caption*{\textbf{Extended Data Fig. 5 | Additional temperature dependence of the memory effect.} Results for \textbf{a,} duration tuning and \textbf{b,} relaxation-rate tuning, to complement the amplitude tuning depicted in Fig.~\ref{fig:temps}. A similar evolution, of diminishing memory effect at elevated temperatures, is seen for all three types of perturbative stimuli.}
    \label{pic:Tdep_3D}
\end{figure}

\begin{figure}[h!]
    \vspace{2cm}
    \begin{center}
    \includegraphics[width=1\linewidth]{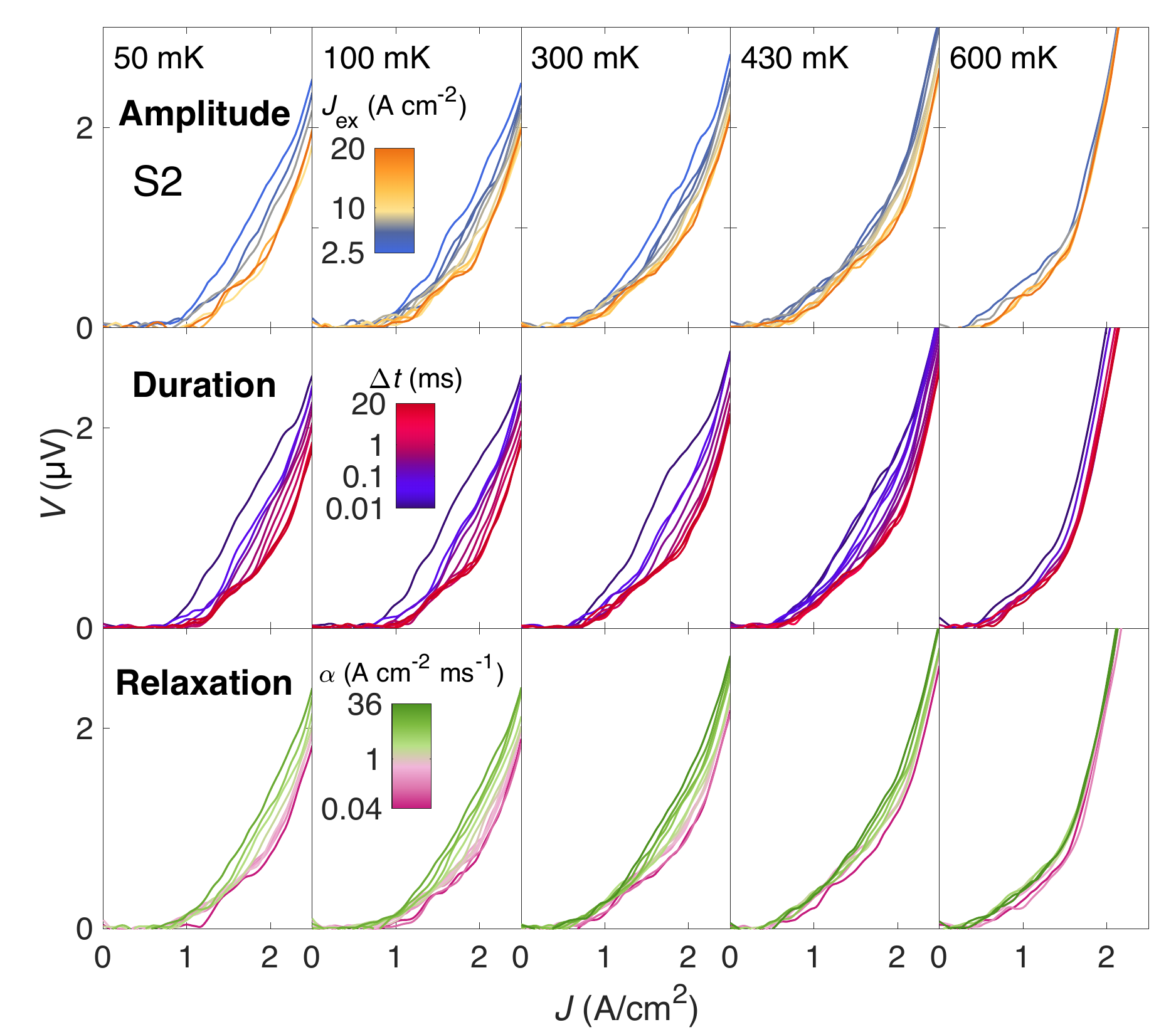} 
    \end{center}
    \vspace{-5pt}
    \caption*{\textbf{Extended Data Fig. 6 | Raw temperature-dependent data.} Hysteretic $J-V$ measurements performed on sample S2 at 14~T at 50~mK, 100~mK, 300~mK, 430~mK and 600~mK for amplitude, duration and relaxation-rate tuning. These data were used to generate the heatplots presented in Fig.~\ref{fig:temps} and Extended Data Fig. 5.}
    \label{pic:3D_data}
\end{figure}

\begin{figure}[h!]
    \vspace{-2cm}
    \begin{center}
    \includegraphics[width=0.7\linewidth]{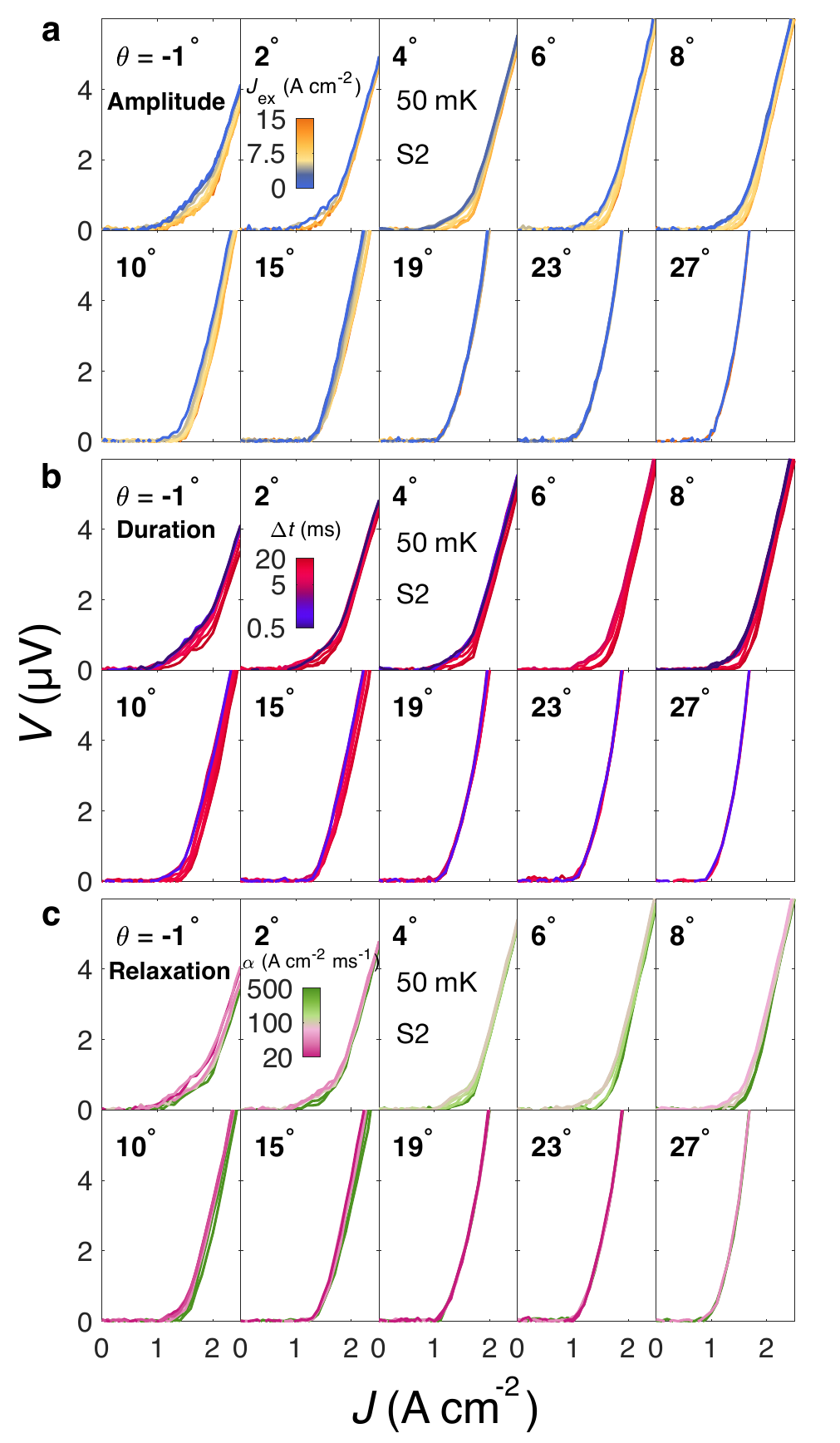} 
    \end{center}
    \vspace{-5pt}
    \caption*{\textbf{Extended Data Fig. 7 | Rotation study of the superconducting memory effect. a,} Amplitude, \textbf{b,} duration and \textbf{c,} relaxation-rate tuning of the memory effect under rotations of $B$ from the $\hat{b}$-axis ($\theta = 0\degree$) towards the $\hat{c}$-axis ($\theta = 90\degree$). $\Delta J$ abruptly diminishes for $\theta \gtrapprox$~19$\degree$.}
    \label{pic:3D_data}
\end{figure}

\begin{figure}[h!]
    \vspace{-3cm}
    \begin{center}
    \includegraphics[width=0.8\linewidth]{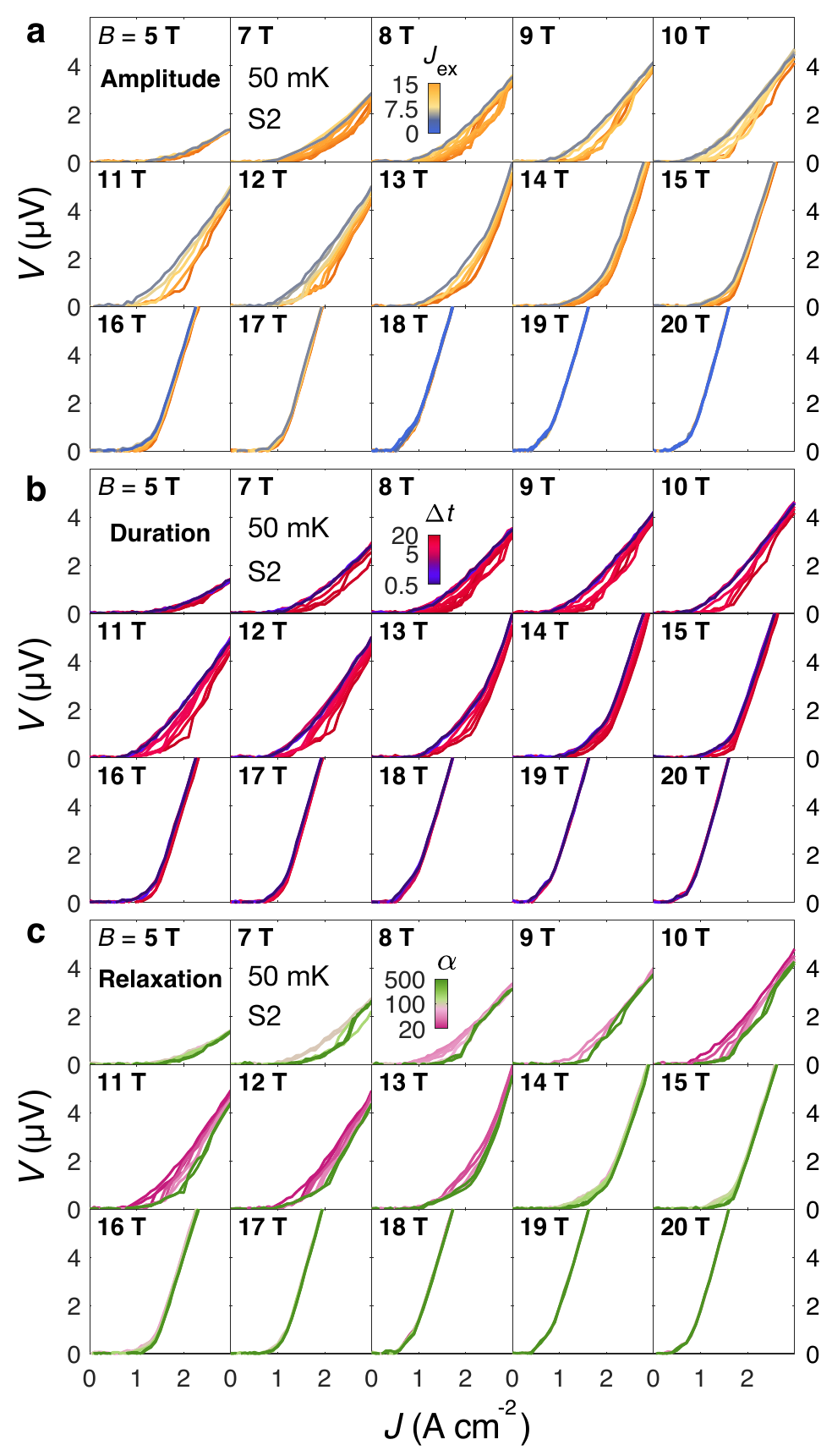} 
    \end{center}
    \vspace{-5pt}
    \caption*{\textbf{Extended Data Fig. 8 | Evolution of the memory effect as a function of magnetic field strength. a,} Amplitude, \textbf{b,} duration and \textbf{c,} relaxation-rate tuning of the memory effect. Measurements were performed on sample S2 for $B$ aligned along the $\hat{b}$-axis.}
    \label{pic:3D_data}
\end{figure}

\begin{figure}[h!]
    \vspace{-3cm}
    \begin{center}
    \includegraphics[width=0.7\linewidth]{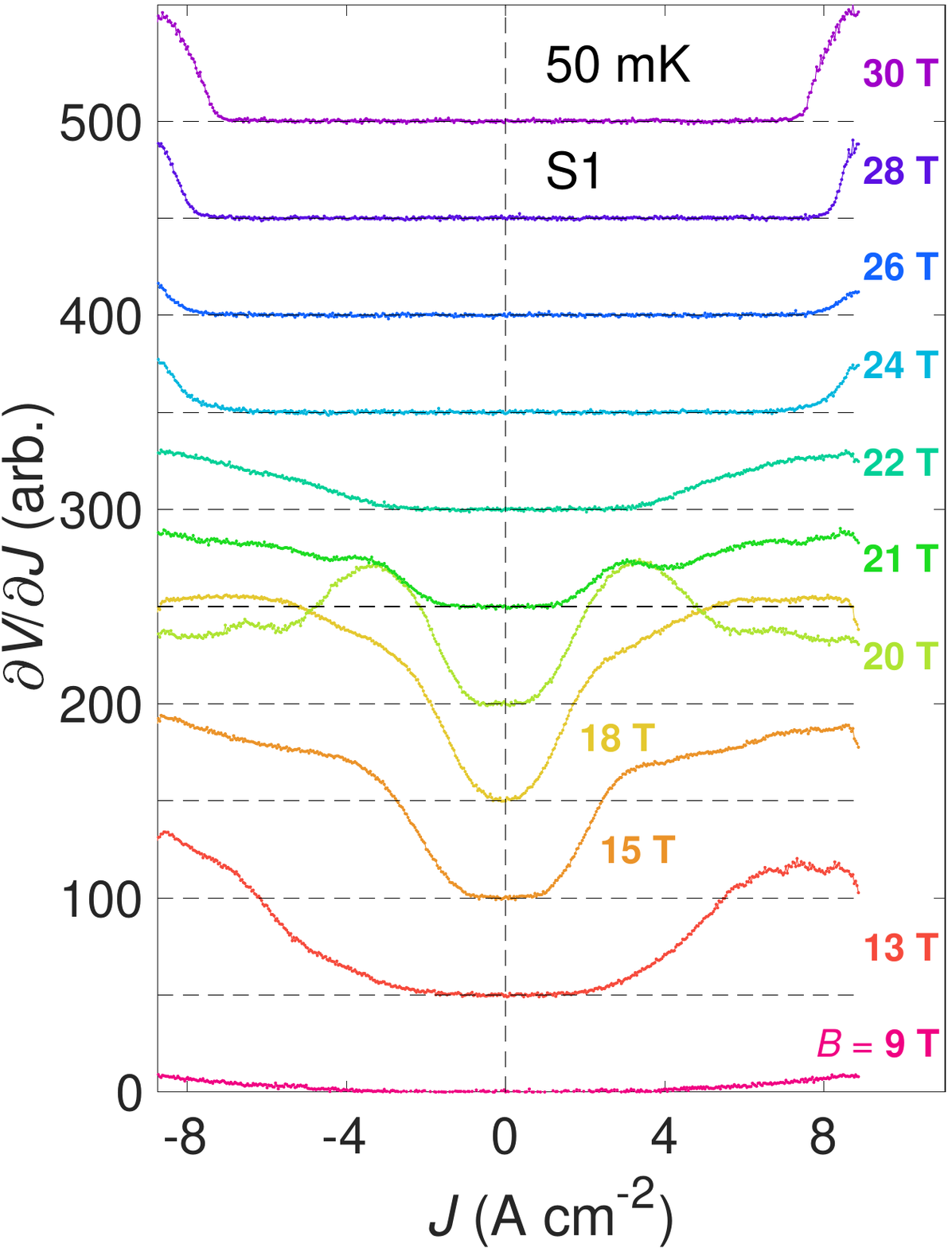} 
    \end{center}
    \vspace{-5pt}
    \caption*{\textbf{Extended Data Fig. 9 | Magnetic field dependence of the slope of $V(J)$ in the equilibrium state.} $\nicefrac{\partial V}{\partial J}$ at incremental field strengths as indicated. $J_c$ is minimal around 20~T, and is actually higher at 28~T than it is at 9~T. Anomalous non-monotonic behaviour is observed in the memory region around 20~T. Note that for this sample (S1) the memory region is located over a higher range of $B$ than for sample S2, likely due to differences in crystalline quality (and hence vortex pinning forces). All measurements were performed at $T = $~50~mK.}
    \label{pic:3D_data}
\end{figure}

\clearpage

\begin{figure}[h!]
    \vspace{-0cm}
    \begin{center}
    \includegraphics[width=.8\linewidth]{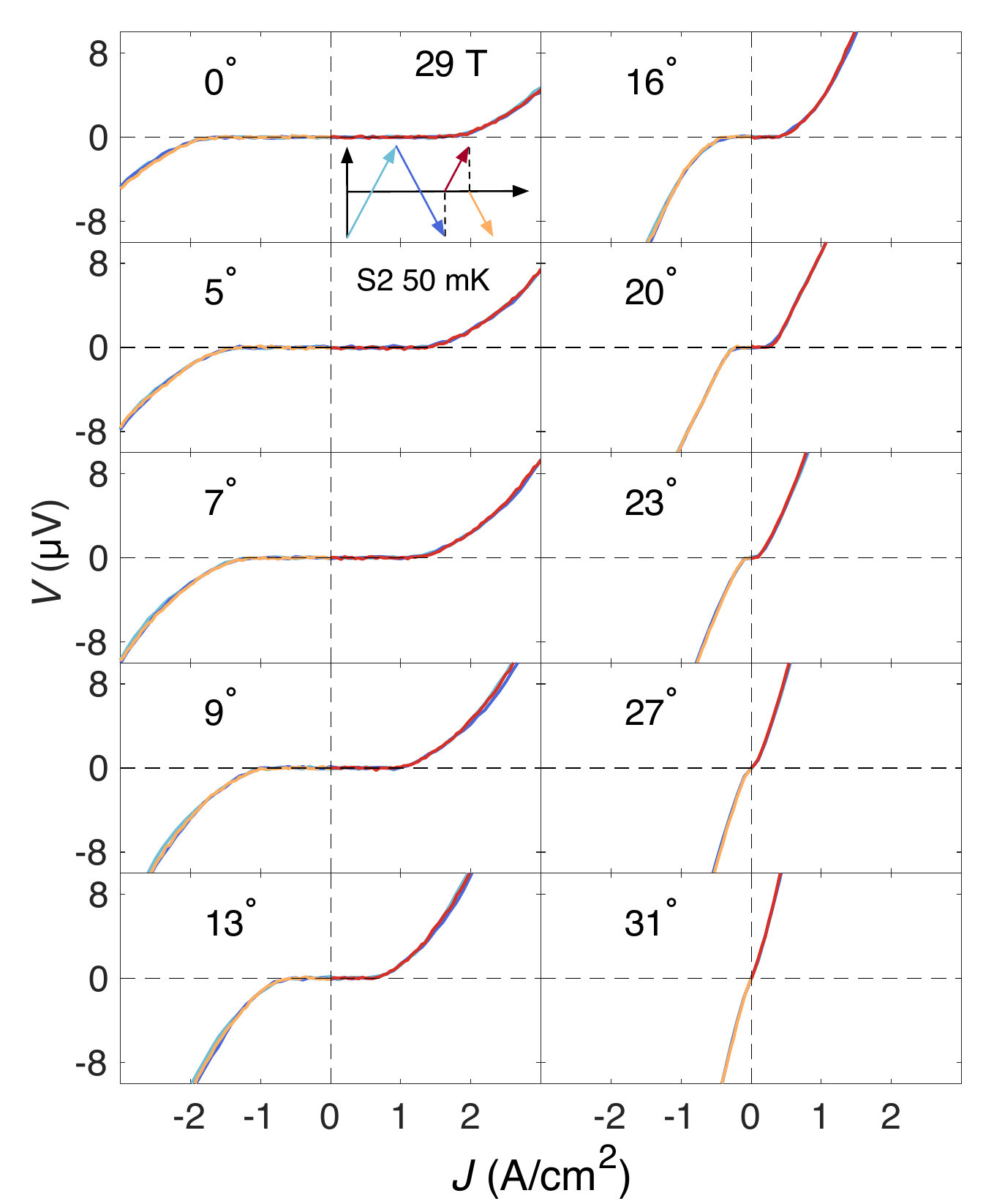} 
    \end{center}
    \vspace{-5pt}
    \caption*{\textbf{Extended Data Fig. 10 | Rotation study showing no memory effect in SC2.} $V(J)$ at $B =$~29~T for rotations of $B$ from the $\hat{b}$-axis ($\theta = 0\degree$) towards the $\hat{c}$-axis ($\theta = 90\degree$). The measurement protocol is specified in the top left inset. No memory effect is discerned at any $\theta$.}
    \label{pic:3D_data}
\end{figure}

\end{document}